\documentstyle[prd,aps,psfig,floats]{revtex}
\def\beq{\begin{equation}}
\def\eeq{\end{equation}}
\def\bea{\begin{eqnarray}}
\def\eea{\end{eqnarray}}
\def\etal{{\it et~al.}}
\def\ie{{\it i.e.}\ }

\def\eq{Eq.\,}
\def\eqs{Eqs.\,}
\def\lam{\lambda}
\def\lamh{\lambda_\chi}
\def\mpl{m_{\text{Pl}}}
\def\mum{\mu_{\text{max}}}
\def\Ncrit{N_{\text{crit}}}
\def\acrit{a_{\text{crit}}}
\def\kcrit{k_{\text{crit}}}
\def\tcrit{t_{\text{crit}}}
\def\af{a_{\text{f}}}
\def\kf{k_{\text{f}}}
\def\tf{t_{\text{f}}}

\def\phif{\phi_{\text{f}}}

\def\xe{x_{\text{e}}}
\def\te{t_{\text{e}}}
\def\cpc{{\cal P}_\chi}
\def\half{\frac{1}{2}}
\def\phidot{\dot{\phi}}
\def\chidot{\dot{\chi}}
\def\flphi2{\langle\delta\phi^2\rangle}
\def\flchi2{\langle\delta\chi^2\rangle}
\def\Phid{\dot{\Phi}}
\def\delphi{\delta\phi}
\def\delchi{\delta\chi}
\def\delphid{\dot{\delta\phi}}
\def\delchid{\dot{\delta\chi}}
\def\phitw{\tilde{\phi}}
\def\delphitw{\delta\tilde{\phi}}
\def\delchitw{\delta\tilde{\chi}}
\def\bra{\langle}
\def\ket{\rangle}
\def\sh{super-Hubble }
\def\ph{preheating }
\def\pr{parametric resonance }
\def\br{backreaction }

\begin{document}
%\bibliographystyle{prsty}
%\bibliographystyle{apsrev}
%\bibliographystyle{h-physrev}
\draft
\twocolumn[\hsize\textwidth\columnwidth\hsize\csname
@twocolumnfalse\endcsname
\title{Backreaction and the Parametric Resonance of Cosmological Fluctuations}
\author{J. P. Zibin$^{1}$, R. Brandenberger$^{1,2}$, and Douglas Scott$^{1}$}
\address{$^1$Department of Physics and Astronomy, 
University of British Columbia, 6224 Agricultural Road, 
Vancouver, BC V6T 1Z1 Canada}
\address{$^2$Physics Department, Brown University, Providence, RI, 02912, USA}
\date{\today}
\maketitle

\begin{abstract}

   \ \ We study the class of two-field inflationary universe 
models $\lam\phi^4/4 + g^2\chi^2\phi^2/2$, in which 
parametric resonance during the initial 
stages of reheating can lead to an exponential amplification of the 
amplitude of cosmological fluctuations.  Employing both analytical 
arguments and numerical simulations, we determine the time at which 
backreaction of fluctuations on the background fields shuts off 
the exponential growth, making use of the Hartree approximation, 
and including scalar metric perturbations.  For the case $g^2/\lam=2$, 
we find that the amplitude of fluctuations after \ph will exceed 
the observational upper bound independent of the value of $\lam$, 
unless the duration of inflation is very long.  Cosmological fluctuations 
are acceptably small for $g^2/\lam\geq8$.  We also find that 
the addition of $\chi$-field self-interaction can limit the growth 
of fluctuations, and in the negative-coupling case the system can 
become effectively single-field, removing the resonance.

\end{abstract}

\vspace{0.35cm}
]

%\tableofcontents

\section{Introduction}

   It has long been realized that reheating is a crucial part 
of the inflationary scenario.  During reheating the large energy 
density contained within the coherently oscillating inflaton 
field is converted into particle excitations of whatever fields 
are coupled to the inflaton, vastly increasing the temperature 
and entropy density and setting the stage for the standard big 
bang phase.  If inflation is ever to be a useful picture for 
describing the early universe, then it is essential to understand the 
details of how the vacuum energy is transformed into familiar particles.

   Reheating can occur very efficiently through the process of 
parametric resonance \cite{earlyph,kls97,stb95}.  Field modes 
within certain resonance bands in $k$-space grow exponentially with 
time, defining the ``preheating'' era.  The possibility of resonant 
growth of linear scalar {\it metric} perturbations was first studied in 
\cite{earlymet}.  Recently it has been argued that the resonance 
of scalar metric perturbations 
can extend to $k\ll aH$, \ie that \sh perturbations can be 
amplified \cite{bkm99,btkm99}.  This opens up the possibility of 
new observational consequences, since the scales relevant to the cosmic
microwave background and large-scale structure are much 
larger than the Hubble radius during preheating.  The importance 
of the gauge-invariant formalism for cosmological perturbations 
\cite{mfb92} and the study of the ``traditionally conserved'' 
Bardeen parameter $\zeta$ was emphasized in \cite{fb99}, where it 
was found that simple single-field chaotic inflation models do not 
exhibit \sh growth beyond what is expected in the absence 
of parametric resonance. The absence of parametric amplification of
super-Hubble modes in these single field models was shown to hold 
in a full nonlinear treatment \cite{pe}, and a general no-go theorem
in these models was suggested in \cite{lmz00}.

   For the first model which was claimed to exhibit growth of 
\sh metric perturbations beyond that of the usual theory 
of reheating \cite{btkm99} (see also \cite{tn98}), it was soon realized 
that the growth was unimportant since it followed a period of exponential 
{\it damping} during inflation \cite{i99,js99,llmw99}.  This damping of 
\sh modes arises because the field perturbations which are 
amplified during preheating have an effective mass greater than the 
Hubble parameter during inflation.  This results in a very ``blue'' 
power spectrum at the end of inflation, with a severe deficit at the 
largest scales \cite{llmw99}.  The relatively plentiful small-scale modes 
can also grow resonantly during preheating.  Thus the 
end of \pr occurs when the \br of the dominant small-scale modes 
becomes important, and the cosmological-scale modes are still negligible.  
An obvious class of models to study, then, consists of those with 
small masses during inflation and strong \sh resonance 
\cite{bgmk00,fb00}.  A simple example was provided by Bassett and 
Viniegra \cite{bv99}, namely that of a massless self-coupled inflaton 
$\phi$ coupled to another scalar field $\chi$, \ie a model with potential
\beq
V(\phi,\chi) = \frac{\lam}{4}\phi^4 + \frac{g^2}{2}\phi^2\chi^2.
\label{bv}
\eeq
This model has been studied in detail, but in the absence of 
metric perturbations, by Greene \etal \cite{gkls97}, who found that 
the model contains a strong resonance band for $\chi$ fluctuations 
which extends to $k=0$ for the choice $g^2=2\lam$.  Bassett and 
Viniegra \cite{bv99} found that super-Hubble metric perturbations 
are resonantly amplified as well in this model (see also \cite{fb00}).

   To date, however, none of the linearized analyses of 
parametric amplification of super-Hubble-scale metric fluctuations 
in the model (\ref{bv}) has included the effects of backreaction on 
the evolution of the fluctuations.  The 
backreaction of the growing modes on the background fields is 
expected to shut the growth down at some point, but exactly when?  
Backreaction is also the only hope to make models which exhibit 
parametric amplification of super-Hubble cosmological 
perturbations compatible with the Cosmic Background Explorer (COBE) 
normalization \cite{blw96}.

   In this paper we will investigate the effects of backreaction on the 
growth of matter and metric fluctuations using the 
Bassett and Viniegra model (\ref{bv}) as our toy model.  We will 
study the growth of scalar field and scalar metric perturbations, 
including the effect of backreaction in the Hartree approximation.  
We carefully treat the evolution during inflation, which can be 
very important for \sh scales.  
We will compare the large-scale normalization predicted for 
this model with the COBE value, and find that, although backreaction is 
crucial in limiting the growth of the fluctuations, the final 
amplitude is larger than allowed by the COBE normalization 
(for supersymmetry-motivated coupling constant values), unless the 
period of inflation is very long.  Note that the final amplitude 
of fluctuations in our model is independent of the scalar field coupling 
constant (unlike what happens without parametric resonance effects), 
but that it may depend on the duration of the inflationary period.  We also 
extend the model to study the effect of $\chi$-field self-coupling, 
which can be important in limiting the growth of fluctuations.

\section{Model and Linearized Dynamics}
\subsection{Equations of motion and analytical theory}
\label{linanal}

   Our model is the two-real-scalar-field, gravitationally minimally 
coupled model specified by the Lagrangian density
\beq
{\cal L} = \sqrt{-g}\left(\frac{1}{2}\partial_\mu\phi\partial^\mu\phi
                        + \frac{1}{2}\partial_\mu\chi\partial^\mu\chi
          - {\lam\over4}\phi^4 - {g^2\over2}\phi^2\chi^2\right).
\eeq
The field $\phi$ drives inflation, while $\chi$ is significant only 
after parametric resonance begins, so the inflationary dynamics 
is essentially that of $(\lam/4)\phi^4$ chaotic inflation.  Note
that the behaviour of this system is expected to be robust under 
the addition of a small mass term $m_\phi^2\phi^2$ with 
$m_\phi\ll\sqrt\lam\mpl$ and for the ratio of coupling constants satisfying 
$g/\sqrt\lam<\sqrt\lam\mpl/m_\phi$ \cite{gkls97}.  In particular, this 
will be the case for supersymmetric models, which motivate the 
choice $g^2=2\lam$ \cite{bks99}.  In addition, we will show that large 
values of $g/\sqrt\lam$ are in fact inconsistent with the significant 
amplification of \sh modes.  On the other hand, 
for $g/\sqrt\lam>\sqrt\lam\mpl/m_\phi$, the theory of ``stochastic 
resonance'' for a massive inflaton may need to be applied \cite{kls97}.

   It is traditional to separate the inflaton field into a 
homogeneous, ``classical'' background $\phi(t)$ and a perturbation 
$\delphi(x,t)$, which begins as sub-Hubble quantum vacuum 
fluctuations early in inflation.  It will be useful to make a 
similar separation for the field $\chi$, although we have no 
reason to expect a non-zero initial homogeneous $\chi$ component 
in this model.  The equations of motion for 
the homogeneous parts of the inflaton and $\chi$ fields are the 
Klein-Gordon equations,
\beq
\ddot{\phi} + 3H\phidot + \lam\phi^3 + g^2\chi^2\phi = 0,
\label{kg}
\eeq
\beq
\ddot{\chi} + 3H\chidot + g^2\phi^2\chi = 0,
\label{kgchi}
\eeq
with Hubble parameter $H=\dot{a}/a$, where $a$ is the scale factor.  
To complete the background dynamics 
we must specify the evolution of the background spacetime metric.  
We assume a spatially flat Friedmann-Robertson-Walker universe, 
for which the 0-0 Einstein equation gives the Friedmann equation
\beq
H^2 = \frac{8\pi}{3\mpl^2}\left[\half\phidot^2 + \half\chidot^2
    + V(\phi,\chi)\right].
\label{fr}
\eeq

   For $\phi\gtrsim\mpl$, the universe undergoes slow-roll inflation, 
with $H$ approximately constant and the scale factor $a$ 
increasing approximately exponentially 
with time.  As slow-roll ends, the ``damping'' term $3H\phidot$ becomes 
less important in \eq(\ref{kg}) and the field begins to oscillate 
about $\phi=0$.  This marks the start of the preheating period.  
Averaged over several oscillations, the equation of state (in the 
absence of backreaction) is very nearly that of a radiation-dominated 
universe \cite{stb95}, and the amplitude of the inflaton's 
oscillations decays as $a^{-1}$.  This is a consequence of the (near) 
conformal invariance of this massless model, which considerably 
simplifies the treatment of \pr as compared with the massive case 
\cite{gkls97,kls97}.  

   In writing the linearized equations of motion for perturbations about 
the background, we will use the longitudinal gauge.  For this model 
the metric can be written \cite{mfb92}
\beq
ds^2 = (1-2\Phi)dt^2 - a^2(t)(1+2\Phi)dx_idx^i,
\eeq
so scalar metric perturbations are described by the single variable 
$\Phi$.  The momentum-space first-order perturbed Einstein and 
Klein-Gordon equations are
\bea
3H\Phid &+& \left( \frac{k^2}{a^2} + 3H^2 \right) \Phi \nonumber\\
        &=& -\frac{4\pi}{\mpl^2} \sum_i \left( \phidot_i\delphid_i
            - \Phi\phidot_i^2 + V_{,i}\delphi_i \right),
\label{perei1}
\eea
\beq
\Phid + H\Phi = \frac{4\pi}{\mpl^2} \sum_i \phidot_i\delphi_i,
\label{perei2}
\eeq
\beq
\ddot{\delphi_i} + 3H\delphid_i + \frac{k^2}{a^2}\delphi_i
 + \sum_j V_{,ij}\delphi_j = 4\Phid\phidot_i - 2V_{,i}\Phi,
\label{perkg}
\eeq
where $\phi_1\equiv\phi$, $\phi_2\equiv\chi$, $V_{,i}\equiv
\partial V/\partial\phi_i$, and comoving momentum $k$ subscripts 
have been suppressed for clarity.  
Equations (\ref{perei1}) and (\ref{perei2}) can be combined to give 
\beq
\Phi = \frac{\sum_i \left( \phidot_i\delphid_i + 3H\phidot_i\delphi_i
                          + V_{,i}\delphi_i \right)}
            {-(\mpl^2/4\pi) (k/a)^2 + \sum_i\phidot^2_i},
\label{constr}
\eeq
which fixes $\Phi$ once the matter fields are known.

   An important quantity in the study of the linear evolution of 
metric perturbations is the Bardeen parameter \cite{bst83}
\beq
\zeta_k = \Phi_k - \frac{H}{\dot{H}} \left( \Phid_k + H\Phi_k \right),
\label{zeta}
\eeq
which for $k/a \ll H$ and {\em single field} models satisfies the 
``conservation law'' \cite{mfb92}
\beq
(1+w)\dot{\zeta_k} = 0,
\label{zetacons}
\eeq
where $w=P/\rho$ is the equation of state ($\rho$ and $P$ denoting energy 
density and pressure, respectively).  
When $\Phid_k$ can be neglected, \eqs(\ref{zeta}) 
and (\ref{zetacons}) can be combined to give the familiar result 
that the change in $\Phi_k$ on super-Hubble scales over some interval 
of time is determined solely by the change in equation of state.

   In the absence of metric perturbations, the linearized dynamics in the 
model described above is known to exhibit \pr during \ph \cite{gkls97}.  
To see this, it helps to take advantage of the near conformal 
invariance of the model and rewrite the equations in terms of 
conformally scaled fields $\phitw_i \equiv a\phi_i$ and 
a dimensionless conformal time $x \equiv \sqrt\lam\phitw_0\eta$, 
where $\eta=\int dt/a$ and $\phitw_0$ is the amplitude of 
inflaton oscillations at the start of preheating.  Then, with the 
$\chi$ background and metric perturbations set to zero, the inflaton 
background and perturbed field equations (\ref{kg}) and (\ref{perkg}) become
\beq
\phitw'' + {\phitw^3\over\phitw_0^2} = 0,
% - {a''\over a}\phitw = 0,
\eeq
\beq
\delphitw_k'' + \left(\kappa^2
 + 3{\phitw^2\over\phitw_0^2}\right)\delphitw_k = 0,
\label{delphiconf}
\eeq
\beq
\delchitw_k'' + \left(\kappa^2
 + {g^2\over\lam}{\phitw^2\over\phitw_0^2}\right)\delchitw_k=0,
%\label{perkg}
\eeq
where $\kappa^2 \equiv k^2/(\lam\phitw_0^2)$ is a dimensionless 
comoving momentum and primes denote derivatives with respect to 
the scaled conformal time $x$.
Here we have ignored terms proportional to $a''/a$ since \ph is a 
nearly-radiation-dominated phase in this model \cite{stb95}.  
The conformal field 
$\phitw$ then undergoes constant amplitude elliptic cosine 
oscillations, while the perturbation equations are Lam\'{e} 
equations \cite{e55}, which are known to exhibit resonance within certain 
bands in parameter space \cite{gkls97}.  In particular, the $\delchitw_k$ 
equation exhibits strong resonance for a band that includes $k=0$ 
for the supersymmetric point $g^2/\lam=2$, and weak ``narrow'' 
resonance in small-scale bands.  On the other hand, $\delphitw_k$ 
exhibits narrow resonance for a sub-Hubble momentum range, 
independent of the coupling constants.  For resonant modes, the 
growth is a modulated exponential, $\delchitw_k\propto e^{\mu_kx}$, 
with Floquet index $\mu_k$.  If we allow $g^2/\lam$ 
to vary, we find a sequence of $\delchitw_k$ resonance bands 
for $k=0$, centred at $g^2/\lam=2n^2$ with width $2n$, for 
positive integral $n$ \cite{gkls97}.  The Floquet index reaches 
a maximum value of $\mum\simeq0.238$ at the centre of each 
$k=0$ band.

\subsection{Numerical results}
\label{numnobr}

   For our numerical calculations, we were primarily interested in the 
behaviour of cosmological-scale matter and metric modes.  Thus we evolved 
a scale which left the Hubble radius (at time $t_0$) at about $N=50$ 
$e$-folds before the end of inflation.  For $(\lam/4)\phi^4$ models, the 
number of $e$-folds during slow-roll inflation after initial time $t_0$ is 
\cite{l90}
\beq 
N\simeq\pi\left({\phi(t_0)\over\mpl}\right)^2;
\label{Nsr}
\eeq
thus we used the homogeneous inflaton initial value of $\phi(t_0)=4\mpl$.  
We began the calculations with the modes still somewhat inside the 
Hubble radius, so the initial conditions for the matter field 
fluctuations were simply given by the conformal vacuum state
\beq
\delphi_{ik}(t_0) = 
   {1\over a^{3/2}(t_0)} \left({1\over{2\omega_i(t_0)}}\right)^{1/2},
\label{ic}
\eeq
\beq
\delphid_{ik}(t_0) = -i\omega_i(t_0)\delphi_i(t_0),
\label{icdot}
\eeq
with $\omega_\phi^2(t) = (k/a)^2 + 3\lam\phi^2 + g^2\chi^2$ and 
$\omega_\chi^2(t) = (k/a)^2 + g^2\phi^2$.  Physically, the $a^{-3/2}$ 
dependence arises because particle number densities 
$n_k \propto |\delphi_{ik}|^2$ must decay like $a^{-3}$ in the 
massive, adiabatic regime.  The initial metric perturbations 
were then determined by \eq(\ref{constr}).

\begin{figure}
%\centerline{\psfig{figure=/home/zibin/ph/paper/fig1.ps,height=8cm}}
\centerline{\psfig{figure=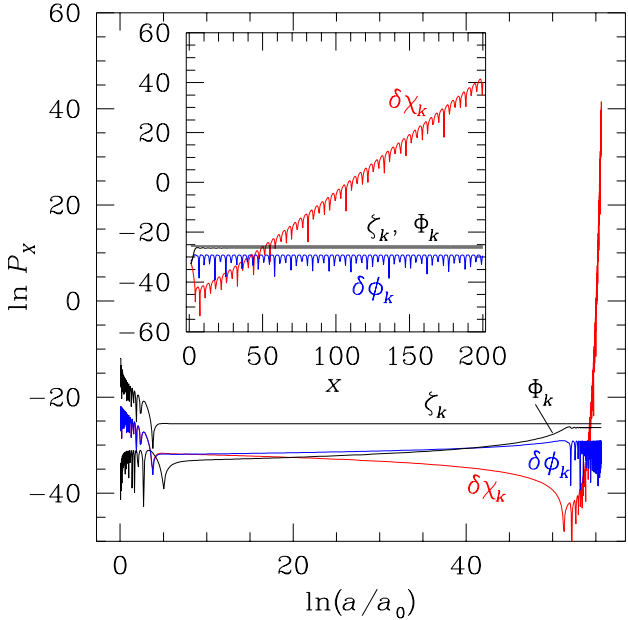,height=8.7cm}}
\caption{Numerical simulation of linear cosmological-scale 
perturbations in the two field model described in the text, 
in the absence of backreaction.  Plotted are the logarithms 
of the power spectra ${\cal P}_X(k) = (k^3/2\pi^2) |X_k|^2$, 
for $X_k = \delchi_k$, $\delphi_k$, $\Phi_k$, and $\zeta_k$, 
and using $\mpl=1$, $g^2/\lam=2$, $\lam=10^{-14}$,
and zero $\chi$ background.  The main figure 
shows the evolution during inflation, while the inset details 
the behaviour during preheating, using the rescaled conformal time 
$x$.  This particular comoving scale leaves the Hubble 
radius approximately $5$ $e$-folds after the start of the simulation.}
\label{nobr}
\end{figure}

   To illustrate the dynamics in the absence of backreaction, we 
numerically integrated the coupled set of background equations 
(\ref{kg}) and (\ref{fr}) and perturbation equations (\ref{perei2}) 
and (\ref{perkg}) using the initial conditions described above, 
and for $g^2/\lam=2$, $\lam=10^{-14}$, and a zero $\chi$ background.  We 
used the constraint \eq(\ref{constr}) as well as the conservation 
equation (\ref{zetacons}) to check the accuracy of the 
calculations.  In Fig. \ref{nobr} we display the evolution of our 
cosmological modes, together with the Bardeen parameter $\zeta_k$, 
during inflation and preheating.  For each of the perturbations 
$X_k = \delchi_k$, $\delphi_k$, $\Phi_k$, and $\zeta_k$ we plot the 
power spectrum \cite{ll93}
\beq
{\cal P}_X(k) = {k^3\over2\pi^2} |X_k|^2,
\label{power}
\eeq
rather than the mode amplitudes, to facilitate comparison with the 
COBE measured normalization which gives ${\cal P}_\Phi\sim10^{-10}$ 
\cite{blw96}.
%The power spectrum ${\cal P}_X(k)$ is a measure of the 
%squared amplitude per logarithmic comoving wave-vector range at $k$.

   Figure \ref{nobr} shows how the modes begin early in inflation 
as sub-Hubble oscillations, and 
become ``frozen in'' after they exit the Hubble radius.  Note that 
the $\delchi_k$ fluctuation experiences some damping late in inflation, 
when its effective mass squared $g^2\phi^2$ becomes somewhat greater than 
$H^2$, which decreases like $\phi^4$ in the slow-roll approximation.  
The inflaton perturbation $\delphi_k$, however, stays roughly constant 
during inflation even though its effective mass is comparable to 
that of the $\delchi_k$ mode.  This is because of the coupling between 
$\delphi_k$ and $\Phi_k$ in the linearized perturbation equations.  
We can also observe a growth of $\Phi_k$ between the time the mode
exits the Hubble radius and the beginning of preheating, by a factor of 
approximately $20$, in good agreement with the growth predicted 
from the ``conservation law'' \eq(\ref{zetacons}).  Also note that after 
this small growth stage the cosmological-scale metric power spectrum ends up 
close to the $10^{-10} (\sim e^{-23})$ level, as the standard theory predicts 
for $\lam\simeq10^{-14}$ in the absence of parametric resonance 
\cite{mfb92}.  During \ph we observe exponential growth of 
$\delchi_k$ while the \sh $\delphi_k$ mode does not grow, as expected 
from the analytical theory.  
$\Phi_k$ and $\zeta_k$ remain constant, since according to 
\eq(\ref{perei2}) the metric perturbations couple only to $\delphi_k$ 
in the absence of a $\chi$ background, at linear level \cite{fb00}.

\begin{figure}
%\centerline{\psfig{figure=/home/zibin/ph/paper/fig2.ps,height=8cm}}
\centerline{\psfig{figure=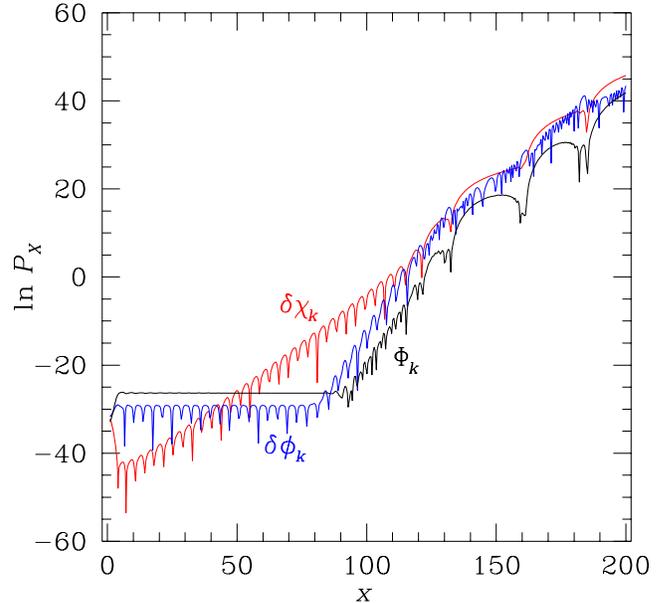,height=8.7cm}}
\caption{Simulation of linear cosmological-scale perturbations 
in the absence of backreaction as in Fig. \ref{nobr}, but with a non-zero 
initial background $\chi(t_0)=10^{-10}\mpl$.  Plotted are the logarithms 
of the power spectra ${\cal P}_X(k) = (k^3/2\pi^2) |X_k|^2$, 
for $X_k = \delchi_k$, $\delphi_k$, and $\Phi_k$, and using $\mpl=1$, 
$g^2/\lam=2$, and $\lam=10^{-14}$.  Coupling through the $\chi$ background 
causes $\delphi_k$ and $\Phi_k$ to grow initially at twice the rate of 
$\delchi_k$.}
\label{nobrbg}
\end{figure}

   To observe the effect of including a non-zero homogeneous $\chi$ 
background, we repeated the above calculation with an initial value 
of $\chi(t_0)=10^{-10}\mpl$ (this value illustrates well the various 
stages of evolution).  Figure \ref{nobrbg} indicates that 
$\delchi_k$ grows as before, but $\delphi_k$ and $\Phi_k$ now grow 
initially with twice the Floquet index of $\delchi_k$.  This is the result 
of the driving term $2g^2\phi\chi\delchi_k$ in the equation of 
motion for $\delphi_k$, \eq(\ref{perkg}), which contains two factors 
growing like $e^{\mum x}$ (clearly the evolution of the background 
$\chi$ will be essentially the same as that of $\delchi_k$ for 
$k\ll aH$).  Once the background $\chi$ 
field becomes comparable to the inflaton background, all the 
perturbations synchronize and grow at the same rate.  We will discuss 
the significance of the homogeneous $\chi$ field in relation to the 
nonlinear evolution of the fields in the next section.

\section{Backreaction}
\subsection{Equations of motion}

   The linearized equations in the previous section describe unbounded 
growth of perturbations during resonance.  In reality this growth 
must of course stop at some point, namely when the 
perturbed field values are on the order of the background values.  
A full nonlinear simulation will include this effect automatically, 
but approximation methods can alleviate the computational costs significantly.  
A common approach to approximate this \br on the background and 
perturbation evolution is to include Hartree terms in the equations 
of motion \cite{earlybr,kls97}.  This entails making the 
replacements $\phi_i \rightarrow \phi_i + \delphi_i$, 
$\delphi_i^2 \rightarrow \bra\delphi_i^2\ket$, and 
$\delphi_i^3 \rightarrow 3\bra\delphi_i^2\ket\delphi_i$.  In this 
approximation, the background equations (\ref{kg}) -- (\ref{fr}) become
\bea
H^2 &=& \frac{8\pi}{3\mpl^2}\Biggl[V(\phi,\chi) \nonumber\\
    & & + {3\over2}\lam\phi^2\flphi2 + {g^2\over2}\phi^2\flchi2
        + {g^2\over2}\chi^2\flphi2 \nonumber\\
    & & \left. + \half\sum_i \left(\phidot_i^2 + \bra\dot{\delphi}_i^2\ket
               + {1\over{a^2}} \bra\nabla\delphi_i^2\ket \right)\right],
\label{frbr}
\eea
\beq
\ddot{\phi} + 3H\phidot + V_{,\phi} + 3\lam\flphi2\phi + g^2\flchi2\phi = 0,
\label{kgbr}
\eeq
\beq
\ddot{\chi} + 3H\chidot + V_{,\chi} + g^2\flphi2\chi = 0.
\eeq
Similarly, the momentum-space linearized field perturbation equations 
(\ref{perkg}) become
%\ddot{\delphi_i} + 3H\delphid_i + \frac{k^2}{a^2}\delphi_i
% + \sum_j V_{,ij}\delphi_j = 4\Phid\phidot_i - 2V_{,i}\Phi,
\bea
\ddot{\delphi}_k &+& 3H\delphid_k + \left(\frac{k^2}{a^2}
 + 3\lam\flphi2 + g^2\flchi2\right)\delphi_k \nonumber\\
   &+& \sum_j V_{,\phi j}\delphi_{jk} = 4\phidot\Phid_k - 2V_{,\phi}\Phi_k,
\label{perkgbr}
\eea
\bea
\ddot{\delchi}_k &+& 3H\delchid_k + \left(\frac{k^2}{a^2}
 + g^2\flphi2\right)\delchi_k \nonumber\\
   &+& \sum_j V_{,\chi j}\delphi_{jk} = 4\chidot\Phid_k - 2V_{,\chi}\Phi_k.
\label{perkgchibr}
\eea
In this approach, the field fluctuations are calculated self-consistently 
from the relations
\beq
\bra\delphi_i^2\ket = {1\over(2\pi)^3} \int d^3k |\delphi_{ik}|^2.
\label{fluct}
\eeq
In practice, the resonance band will provide a natural ultraviolet cutoff.

   The Hartree terms approximate the full nonlinear dynamics of the 
fields.  To illustrate what this approximation entails, we may 
consider the exact dynamics of the fields, treating 
the Klein-Gordon equation as a classical field equation.  This 
should be a good approximation soon after the beginning of 
the resonance stage, since 
occupation numbers will grow exponentially with time \cite{kls97,gkls97}.
As an example, consider the exact evolution equation for $\delphi$ 
in position space, obtained by perturbing the Klein-Gordon equation, 
setting the background $\chi$ to zero, and ignoring metric perturbations,
\bea
\ddot{\delphi} + 3H\delphid &+& {1\over a^2}\nabla^2\delphi
 + 3\lam\phi^2\delphi + 3\lam\phi\delphi^2 + \lam\delphi^3 \nonumber\\
               &+& g^2\delchi^2\phi + g^2\delchi^2\delphi = 0.
\label{delphieom}
\eea
The terms in this equation describing the interaction between 
the $\phi$ and $\chi$ fields become in momentum space
\bea
& & {g^2\phi\over(2\pi)^{3/2}} \int d^3k'
     \delchi_{{\bf k}'}\delchi_{{\bf k}-{\bf k}'} \nonumber\\
&+& {g^2\over(2\pi)^3} \int d^3k'd^3k''
     \delchi_{{\bf k}'}\delchi_{{\bf k}''}\delphi_{{\bf k}-{\bf k}'-{\bf k}''}.
\label{nlbr}
\eea
Thus the Hartree term $g^2\flchi2\delphi_k$ in \eq(\ref{perkgbr}) 
corresponds to the second term in expression (\ref{nlbr}), restricted 
to ${\bf k}''=-{\bf k}'$.  Physically, this means that only scattering events 
which do not change the $\delphi_k$ momentum are included in the 
Hartree approximation, and ``rescattering'' events are ignored.

   It is important to notice that the first 
term in (\ref{nlbr}), which scatters particles from the homogeneous 
inflaton background into mode $\delphi_k$, could be larger than 
the Hartree term since initially $\vert \phi \vert > \vert \delta \phi \vert$, 
unless the first term vanishes upon averaging (integrating) over the 
entire phase space of contributing terms (which is what is 
assumed in the Hartree approximation).  If it does not vanish, the 
first term in (\ref{nlbr}) will act as a driving term for the 
$\delphi$ modes in (\ref{delphieom}).  Since some $\delta \chi$ modes 
experience parametric amplification with Floquet exponent $\mu$, this 
term will lead to an important second-order effect, namely the growth 
of $\delphi$ as $e^{2\mu x}$.  This effect is left out in the 
Hartree approximation.  Because the metric perturbations are 
coupled to $\delphi$ through 
\eq(\ref{perei2}), we also expect that, with the homogeneous $\chi$ 
set to zero, the Hartree approximation will miss the 
corresponding growth of $\Phi$.  Note, however, that by including the $\chi$ 
background term $2g^2\chi\phi\delchi$ in (\ref{delphieom}), and 
setting $\chi^2\sim\flchi2$, we can approximate the effect of the important 
first term in (\ref{nlbr}), as we saw in Fig. \ref{nobrbg}.

   Just as with the scalar fields, metric fluctuations may grow 
rapidly in our model.  We can account for the \br of metric 
perturbations through the effective energy-momentum tensor 
formalism of Abramo \etal \cite{abm}.  This involves expanding 
the Einstein equations to second order in the perturbations and 
taking the spatial average to obtain effective background equations.  
In our case, the metric and inflaton equations (\ref{frbr}) and 
(\ref{kgbr}) become, with background $\chi$ set to zero,
\bea
H^2 &=& \frac{8\pi}{3\mpl^2}\left[\half\phidot^2 + V(\phi)
        + {3\over2}\lam\phi^2\flphi2+{g^2\over2}\phi^2\flchi2\right.\nonumber\\
    & & + \left.\half\sum_i \left(\bra\dot{\delphi}_i^2\ket
        + {1\over{a^2}} \bra(\nabla\delphi_i)^2\ket \right)
        + 2\lam\phi^3\bra\Phi\delphi\ket \right] \nonumber\\
    & & + 4H\bra\Phi\Phid\ket - \bra\Phid^2\ket
        + {3\over{a^2}} \bra(\nabla\Phi)^2\ket,
\label{frmbr}
\eea
\bea
(\ddot{\phi} &+& 3H\phidot)(1 + 4\bra\Phi^2\ket) + \lam\phi^3
                 + 3\lam\flphi2\phi \nonumber\\
             &+& g^2\flchi2\phi - 2\bra\Phi\ddot{\delphi}\ket
                 - 4\bra\Phid\delphid\ket \nonumber\\
             &-& 6H\bra\Phi\delphid\ket + 4\phidot\bra\Phid\Phi\ket
                 - {2\over{a^2}} \bra\Phi\nabla^2\delphi\ket = 0.
\label{kgmbr}
\eea

\subsection{Analytical estimates}
\subsubsection{Evolution of perturbations during inflation}

   Perturbations will grow during \pr until backreaction becomes 
important.  We can analytically estimate the amount of growth 
by estimating the time at which the Hartree term $g^2\flchi2$ 
is of the order of the background $\lam\phi^2$ ({\it cf } 
\eq(\ref{perkgbr})).  Note that in the absence of metric fluctuations, 
such an estimate should be accurate, at least for $g^2/\lam\sim1$, 
as nonlinear lattice simulations indicate \cite{pr97,kt}.  In 
addition, we expect the matter sector to dominate the dynamics.  In 
order to estimate the variance $\flchi2$, we will need to calculate 
the evolution of $\delchi_k$ modes, starting from the adiabatic 
vacuum inside the Hubble radius, continuing through inflation, and 
finally through preheating.  The evolution during inflation is 
quite complicated, and will 
have a crucial effect on the final variances, so we will describe the 
inflationary stage in some detail.  We consider general 
values of $g^2/\lam$, rather than just the supersymmetric point.
%Note that in the absence of metric fluctuations, such an estimate has been shown 
%to be consistent with the results of nonlinear lattice simulations \cite{pr97}.  
%The evolution of resonant modes during 
%preheating is given by the simple growth law $\delchi_k\propto e^{\mu_k x}$ 
%(recall that $x$ is a rescaled conformal time).  

   We will only need to consider the contribution to $\flchi2$ from 
modes which are \sh at the start of preheating.  To see this, first 
note that for $g^2/\lam=2$, the small-scale boundary of the strongest 
(and largest-scale) resonance band is at $k_{\text{max}}/a 
\simeq \sqrt{\lam}\phi_0/2$, where $\phi_0(t)$ is the amplitude of 
inflaton oscillations during preheating \cite{gkls97}.  Next, we can 
use the Friedmann equation (\ref{fr}) to write the Hubble parameter 
in terms of $\phi_0$, giving
\beq
H^2 = {2\pi\over 3\mpl^2}\lam\phi_0^4.
\label{H2sr}
\eeq
(Note that this equation also applies approximately during slow-roll.)  
Using the value $\phi_0=0.2\mpl$, we calculate the ratio 
$aH/k_{\text{max}} \simeq 0.6$ at the start of preheating.  Thus the 
Hubble radius corresponds closely to the smallest resonant scale.  
This result is not very sensitive to $g^2/\lam$ as long as we are near 
the centre of a band, \ie $g^2/\lam=2n^2$, since $k_{\text{max}}$ increases 
only slowly with $g^2/\lam$ in this case \cite{gkls97}.  Also, we can 
ignore the resonance bands at higher $k$ values, since they correspond to 
narrow resonance.
%$\kappa_{\text{max}}=0.5$, for $g^2/\lam=2$ This 
%corresponds to a minimum resonant physical scale of 
%This is not surprising, since $k_{\text{max}}/a$ corresponds to the 
%inflaton frequency, which is necessarily of order $H$ at the end of inflation.  

   To estimate the evolution of $\delchi_k$ on \sh scales during inflation, 
we can ignore terms containing the background $\chi$ as well as the spatial 
gradient term in \eq(\ref{perkg}), resulting in a damped harmonic 
oscillator equation with time-dependent coefficients,
\beq
\ddot{\delchi}_k + 3H\delchid_k + g^2\phi^2\delchi_k = 0.
%\label{delchi}
\eeq
During slow-roll, we can use the adiabatic approximation to find 
solutions to this equation, since $|\dot{H}| \ll H^2$.  Thus for 
$g^2\phi^2 > (3H/2)^2$ we have underdamped oscillations with damping envelope
\beq
\delchi_k \propto \exp\left[-\int (3H/2) dt\right] = a^{-3/2}.
\label{udamp}
\eeq
For $g^2\phi^2 < (3H/2)^2$, we have the overdamped case with two 
decaying modes.  Ignoring the more rapidly decaying mode, we obtain
\beq
\delchi_k \propto \exp\left[-\int \left(3H/2 -
                            \sqrt{9H^2/4 - g^2\phi^2}\right) dt\right].
\label{odamp}
\eeq
In this case, the fluctuations are very slowly decaying in the massless limit 
$g^2\phi^2 \ll (3H/2)^2$, while they approach the $a^{-3/2}$ decay
as $g^2\phi^2 \rightarrow (3H/2)^2$.

   During slow-roll we have $H^2 \propto \phi^4$ (see \eq(\ref{H2sr})),
so that $H^2$ decreases more rapidly than $g^2\phi^2$, and there 
is a transition between the over- and underdamped stages.  The two 
types of behaviour are separated by the critically damped 
case, $g^2\phi^2 = (3H/2)^2$.  Using \eqs(\ref{H2sr}) and (\ref{Nsr}), 
we can write this critical damping condition in terms of the number 
of $e$-folds after critical damping, $\Ncrit$, as
\beq
\Ncrit = \ln\left({\af\over\acrit}\right) = {2\over3}{g^2\over\lam}
       \simeq \ln\left({\kf\over\kcrit}\right),
\label{Ncrit}
\eeq
where subscript ``f'' refers to the end of inflation and ``crit'' to the 
time of critical damping.  Wavevectors $\kcrit$ and $\kf$ leave the Hubble 
radius at $\tcrit$ and $\tf$, respectively.  We see that as $g^2/\lam$ 
increases, cosmological scales are damped like $a^{-3/2}$ during a 
greater and greater part of inflation.  We thus expect that for large 
enough $g^2/\lam$, the backreaction of the smaller-scale modes will 
terminate \pr when cosmological-scale $\delchi_k$ modes are still 
greatly suppressed.  In other words, there 
will be a maximum value of $g^2/\lam$ for which there is significant 
amplification of \sh $\delchi_k$ perturbations, as anticipated in \cite{bv99}.
%as described in the Introduction for the case of 
%a massive inflaton model \cite{i99,js99,llmw99}.
%if a mode leaves the Hubble radius during the overdamped stage.  
%   To make this prediction quantitative, we will now calculate the 
%field variance $\flchi2$.  

   We first consider the evolution of the modes which 
leave the Hubble radius after $\tcrit$, \ie $k > \kcrit$ (but which 
are still \sh at the end of inflation, $k < \kf$).  These modes 
are effectively massive during inflation, and hence we can simply 
use the adiabatic vacuum state, \eq(\ref{ic}), which for $k \ll aH$ gives
\beq
|\delchi_k(\tf)|^2 = {1\over 2\af^3 g\phif}.
\label{smscale}
\eeq
Note that if we define the spectral index $n$ through 
${\cal P}_\chi(k) \propto k^{n-1}$ \cite{ll93}, then for this part of 
the spectrum we have $n=4$, an extreme blue tilt.

   Next, we will calculate the evolution of modes 
which leave the Hubble radius before $\tcrit$, \ie modes with 
$k < \kcrit$.  In this case, the modes are approximately massless when 
they exit the Hubble radius ($g^2\phi^2 < (3H/2)^2$ for $t<\tcrit$), so 
we can use the standard result for a massless inflaton \cite{kt90},
\beq
|\delchi_k(t_k)|^2 = {H^2(t_k) \over 2k^3},
\eeq
where $t_k$ is the time that mode $\delchi_k$ exits the Hubble radius.  
We now must use \eq(\ref{odamp}) to evolve the modes during the overdamped 
period, $t_k < t < \tcrit$.  Writing $dt = d\phi/\phidot$, and using 
the slow-roll approximation $\phidot \simeq -V_{,\phi}/3H$, we can 
perform the integral to obtain
\beq
|\delchi_k(\tcrit)|^2 = {H^2(t_k)\over2k^3}e^{-3F(N_k)},
\eeq
where $N_k$ is the number of $e$-folds after time $t_k$ and 
\bea
F(N_k) &\equiv& N_k-\Ncrit - \sqrt{N_k}\sqrt{N_k-\Ncrit} \nonumber\\
&+& \Ncrit \ln\left({\sqrt{N_k} + \sqrt{N_k-\Ncrit} \over \sqrt{\Ncrit}}\right).
\label{Fdef}
\eea
Next we can readily propagate the modes through the underdamped period, 
$\tcrit < t < \tf$, using \eqs(\ref{udamp}) and (\ref{Ncrit}), giving
\beq
|\delchi_k(\tf)|^2 = {H^2(t_k)\over2k^3} e^{-3F(N_k) - 2g^2/\lam}.
\label{fldamp}
%&=& {H^2(t_k) \over 2k^3} \left({\acrit\over\af}\right)^3 \\
\eeq
Since the damping term $F(N_k)$ is positive, we see as expected that 
large-scale modes are strongly damped for large $g^2/\lam$.

   Finally, we can approximate the conformal time dependence of 
all \sh modes during \pr as
\beq
\delchi_k \propto e^{\mum x},
\label{phevol}
\eeq
if we are near the centre of a resonance band.  This is valid since, in 
this case, the Floquet index $\mu_k$ varies only slightly for scales larger 
than a few times the Hubble radius (\ie the smallest resonant scale) 
\cite{gkls97}.

\subsubsection{Variances and total resonant growth}

   Now we can proceed to calculate the field variance, $\flchi2$.  We 
will use \eq(\ref{fluct}), restricting the integral to the resonantly 
growing modes.  We begin with the case $g^2/\lam=2$.  Equation (\ref{Ncrit}) 
tells us that in this case $\Ncrit\simeq4/3$, so that essentially all of the 
evolution during inflation is in the overdamped regime, and we only 
need to consider modes with $k<\kcrit$.  The variance integral will be 
dominated by modes with $N_k\gg\Ncrit$, so we may approximate the damping 
term in \eq(\ref{Fdef}) as
\beq
e^{-3F(N_k)} \simeq \left({\Ncrit\over4N_k}\right)^{g^2/\lam}.
\label{Fapprox}
\eeq
For the current case, $g^2/\lam=2$, we can now combine the expression 
(\ref{fldamp}) with \eqs(\ref{Nsr}), (\ref{H2sr}), (\ref{phevol}), and 
(\ref{Fapprox}) to obtain for the power spectrum on resonant scales at 
the end of \ph
\beq
\cpc(k,\te) = {\lam\mpl^2 \over 54\pi^3} e^{2\mum\xe}.
\eeq
Here $\te$ is the time that the resonance shuts down, and $\xe$ is the 
corresponding scaled conformal time.  As we will see, the important 
thing about this result is that the power spectrum is essentially 
Harrison-Zel'dovich (independent of $k$), with spectral index $n=1$.
%this means that our final result will be independent of the precise 
%magnitude of $\cpc(\tf)$.

   We can next rewrite the variance integral, \eq(\ref{fluct}), in terms 
of the power spectrum as
\beq
\bra\delchi^2(\te)\ket = \int_0^{N_0} dN_k \cpc(k,\te) = N_0 \cpc(\te),
\label{flps}
\eeq
where $N_0 \simeq 50$ is the total number of $e$-folds during inflation.  
Finally, the criterion $g^2\bra\delchi^2(\te)\ket \sim \lam\phi^2(\te)$ 
gives, using the value $\phi(\te) \sim 10^{-2}\mpl$,
\beq
\cpc(\te) \sim 10^{-6}\mpl^2
\label{brval2}
\eeq
for the $\delchi_k$ power spectrum on cosmological scales at the end of 
preheating.  Note that this result used only the $k$-independence 
of the power spectrum (which is a result of the special choice $g^2/\lam=2$), 
and the values of $N_0$ and $\phi(\te)$.  In particular, the result 
is independent of $\lam$, unless, contrary to our implicit assumption, 
$\lam$ is so large that $g^2\bra\delchi^2\ket > \lam\phi^2$ already 
at the start of preheating.  In this case, \eq(\ref{brval2}) will be 
an underestimate.  

   According to the results from Section \ref{numnobr}, we expect 
synchronization of the other fields to $\delchi_k$, so that in 
particular we expect ${\cal P}_\Phi \sim \cpc/\mpl^2$.  Therefore we 
conclude that, for $g^2/\lam=2$, the metric perturbation amplitude 
will indeed be considerably larger than the COBE measured value, 
even including the effect of backreaction.

   Next we will repeat the preceding analysis for the second \sh resonance 
band, at $g^2/\lam=8$.  In this case we have $\Ncrit\simeq5$, so we must 
consider modes that exit the Hubble radius both before and after $\tcrit$.  
For the large-scale modes, $k<\kcrit$, it will be sufficient to place 
an upper limit on the variance.  Using \eq(\ref{fldamp}), but ignoring 
the damping factor $e^{-3F}$, we obtain
\beq
\cpc(k,\tf) < {H^2(\tcrit) \over (2\pi)^2} e^{-2g^2/\lam}
            \simeq 2\times10^{-8}\lam\mpl^2
\eeq
on scales $k<\kcrit$ at the end of inflation.  Thus, using \eq(\ref{flps}), 
the contribution to the variance from modes with $k<\kcrit$ satisfies the 
(probably very conservative) bound
\beq
\bra\delchi^2(\tf)\ket_{k<\kcrit} < 9\times10^{-7}\lam\mpl^2.
\label{lgvar}
\eeq
Next we can use \eq(\ref{smscale}) to calculate the contribution to 
$\flchi2$ from smaller-scale modes with $\kcrit < k < \kf$,
\bea
\bra\delchi^2(\tf)\ket_{\kcrit<k<\kf}
  &=& {1 \over 16\pi^3 \af^3 g\phif} \int_{\kcrit}^{\kf} d^3k \\
  &=& {1\over18} \sqrt{{2\over3\pi}{\lam \over g^2}} {\lam\phif^5\over\mpl^3} \\
 &\simeq& 3\times10^{-6}\lam\mpl^2.
\label{smvar}
\eea
Here we have used $\kcrit^3 \ll \kf^3$ (which follows from \eq(\ref{Ncrit}) 
for $g^2/\lam=8$), the relation $\kf/\af=H(\tf)$, \eq(\ref{H2sr}), and the value 
$\phif=0.2\mpl$.  This value of the small-scale variance exceeds our 
upper limit on the large-scale variance in \eq(\ref{lgvar}), so we can 
ignore the contribution from the large-scale modes, 
$\bra\delchi^2(\tf)\ket_{k<\kcrit}$.
%  Note that the 
%variance is only weakly dependent on $g^2/\lam$ on these small scales.

   Now we can again apply the condition $g^2\bra\delchi^2(\te)\ket 
\sim \lam\phi^2(\te)$, which in this case gives
\beq
e^{2\mum\xe} \sim 4\lam^{-1}.
\eeq
Finally, we can use \eq(\ref{fldamp}) without approximation to calculate 
the cosmological-scale power spectrum at the end of preheating, for the 
case $g^2/\lam=8$,
\bea
\cpc(\te) &=& {H^2(t_0) \over (2\pi)^2} 
              \exp\left[-3F(N_0) - 2{g^2\over\lam} + 2\mum\xe\right] \\
         &\sim& 10^{-14}\mpl^2. \label{brval8}
\eea
In this case the growth stops before the cosmological 
perturbations exceed the COBE value, and thus \pr does not change 
the standard predictions \cite{mfb92} for the size of the fluctuations.  
Therefore, since the damping of \sh $\delchi_k$ modes increases as 
$g^2/\lam$ increases, the standard predictions are not modified for 
all resonance bands beyond the first, \ie for $g^2/\lam \geq 8$.

   To close this section, consider the behaviour of the large-scale variance 
if we suppose that inflation started much earlier than the time that 
cosmological scales left the Hubble radius, \ie let 
$N_0 \rightarrow \infty$.  In this limit we see from the approximation 
\eq(\ref{Fapprox}), the expression $H(t_k) \propto N_k$, and \eq(\ref{fldamp}) 
that the variance becomes
\beq
\bra\delchi^2\ket_{k<\kcrit} \propto \int^{N_0}_{\Ncrit} dN_k N_k^{2-g^2/\lam},
\eeq
which, for $g^2/\lam=2$, is divergent as $N_0 \rightarrow \infty$.  Such 
divergences are well-known in inflationary models \cite{sh98}.  Here the 
divergence suggests that for large enough $N_0$, the growth of 
cosmological-scale modes will stop before they exceed the COBE amplitude, 
due to the large contribution to the variance from super-cosmological 
scales.  Indeed, for $N_0\sim10^6$ (a value not out of the question 
in chaotic inflation \cite{l90}) \eq(\ref{flps}) gives 
$\cpc(\te)\sim10^{-10}\mpl^2$ for $g^2/\lam=2$.  For $g^2/\lam>3$ the 
variance converges, although in this case cosmological-scale modes are 
already supressed by the mechanism described above.
%res only up to coh scale o' inflaton  
%and so our previous results are unchanged.
%Since the initial fluctuations are of order $\lam^{1/2}$, we see that 
%when the growth stops the fluctuations will be of order $N^{-3/2}$, 
%independent of $\lam$.  [meaning/relevance of this IR divergence (start
%of inflation...)]

\subsection{Numerical Results}

   It is straightforward to check our analytical estimates from the 
previous section by numerically integrating the coupled set of 
Hartree approximation evolution equations (\ref{frbr}) -- (\ref{fluct}) 
and metric perturbation equation (\ref{perei2}).  
We now must evolve a set of modes that fill the relevant resonance 
band.  For example, for $g^2/\lam=2$, the first resonance band extends 
from $\kappa=0$ to $\kappa=0.5$ \cite{gkls97}.  Again we begin each 
mode's evolution inside the Hubble radius during inflation, using 
the initial vacuum state, \eqs(\ref{ic}) and (\ref{icdot}).  Each 
mode is incorporated into the calculation shortly before it leaves 
the Hubble radius, so that the spatial gradient terms are never 
too large.  The variances are calculated 
by performing the discretized integrals, \eqs(\ref{fluct}), 
only over the resonance band; thus they are convergent.  Note that 
the variances are calculated {\em simultaneously} with the field 
backgrounds and perturbations.

\begin{figure}
%\centerline{\psfig{figure=/home/zibin/ph/paper/fig3.ps,height=8cm}}
\centerline{\psfig{figure=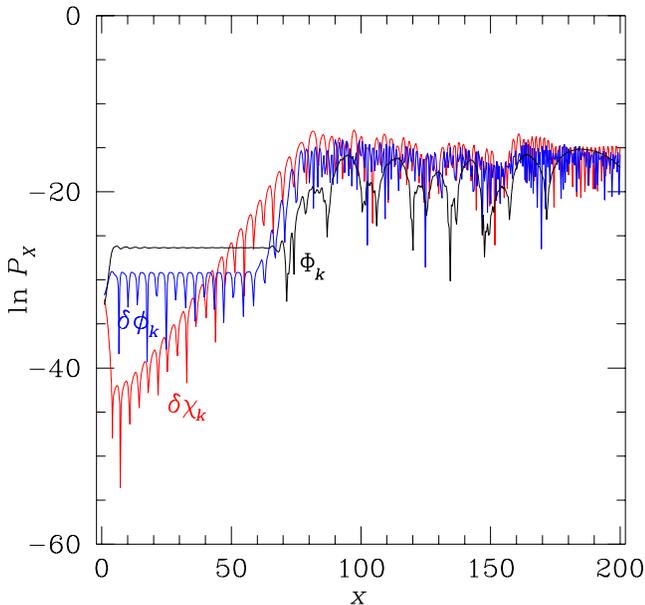,height=8.7cm}}
\caption{Numerical simulation of cosmological-scale perturbations 
with Hartree backreaction terms included and with a non-zero 
initial background, $\chi(t_0)=10^{-6}\mpl$.  Plotted are the 
logarithms of the power spectra ${\cal P}_X(k) = (k^3/2\pi^2) |X_k|^2$, 
for $X_k = \delchi_k$, $\delphi_k$, and $\Phi_k$, 
using $\mpl=1$, $g^2/\lam=2$, and $\lam=10^{-14}$.  Backreaction 
terminates the growth of each field perturbation.}
\label{br}
\end{figure}

   In Fig. \ref{br} we present the evolution of the $\delchi_k$, 
$\delphi_k$, and $\Phi_k$ power spectra on the same cosmological scale 
as was studied in Section II.  All parameters are the same as for 
Fig. \ref{nobrbg}, except here we use for the initial background value 
$\chi(t_0)=10^{-6}\mpl$, which means that during preheating 
$\chi^2\simeq\flchi2$.  The evolution is initially similar to 
that of Fig. \ref{nobrbg}, only here the growth saturates at 
${\cal P}_\chi \sim 3\times10^{-7}\mpl^2$, in good agreement with our 
prediction based on \eq(\ref{brval2}).  Also, as expected, the other 
fields closely follow ${\cal P}_\chi$.  Whereas in the linear 
calculations the Einstein constraint equation (\ref{constr}) was 
satisfied to extremely good accuracy, with the inclusion of backreaction 
${\cal P}_\Phi$ saturates at a factor of roughly $10^3$ higher using 
\eq(\ref{constr}) than the illustrated result, which used \eq(\ref{perei2}).  
Note that a similar result was found in \cite{js99}.  We suspect that
this is a fundamental problem related to our attempt to capture some 
of the nonlinear dynamics with the Hartree approximation.  Regardless 
of which value is used, the cosmological metric perturbations 
considerably exceed the COBE normalisation.  In addition, we find no 
significant difference in the results when backreaction of metric 
perturbations is included using \eqs (\ref{frmbr}) and (\ref{kgmbr}), 
as expected if the matter fields dominate the backreaction.  Thus all 
of our presented results exclude the metric backreaction terms.

\begin{figure}
%\centerline{\psfig{figure=/home/zibin/ph/paper/fig4.ps,height=8cm}}
\centerline{\psfig{figure=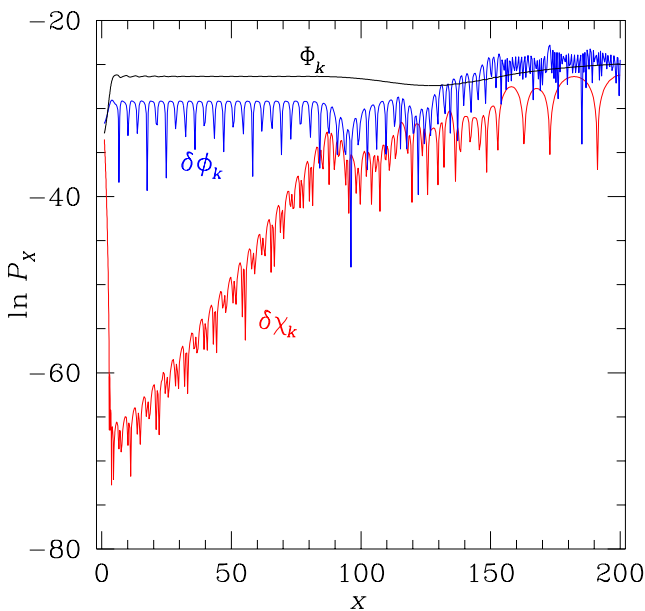,height=8.7cm}}
\caption{Same as Fig. \ref{br}, except for parameters lying in the second 
resonance band, \ie $g^2/\lam=8$, and with $\chi(t_0)=10^{-6}\mpl$ 
and $\lam=10^{-14}$.  Here \br of the small-scale modes terminates the 
growth before the COBE value is exceeded.}
\label{g8}
\end{figure}

   As discussed above, larger values of $g^2/\lam$ result in increased 
damping of $\delchi_k$ on large scales during inflation, and at large enough 
$g^2/\lam$ we expect insignificant amplification of \sh modes.  This 
is illustrated in Fig. \ref{g8}.  Here we examine the second resonance 
band at $g^2/\lam=8$, but use otherwise identical parameters to 
Fig. \ref{br}.  Resonance stops at ${\cal P}_\chi \sim 10^{-14}\mpl^2$, 
consistent with our analytical estimate from \eq(\ref{brval8}), 
and not exceeding the standard predictions for $\lam\simeq10^{-14}$ 
\cite{mfb92}.  Note that the small rise in ${\cal P}_\Phi$ at late 
times should not be trusted, as our Hartree approximation scheme will 
not capture the full nonlinear behaviour.  
For resonance bands at even higher $g^2/\lam$, 
we find extremely suppressed cosmological $\delchi_k$ amplitudes, in 
quantitative agreement with the calculations of the previous section.

\pagebreak

\section{Self-interacting $\chi$ models}
\subsection{Positive coupling}

   We now consider the addition of a quartic self-interaction term 
for the $\chi$ field, so that our potential becomes
\beq
V(\phi,\chi) = \frac{\lam}{4}\phi^4 + \frac{g^2}{2}\phi^2\chi^2
             + \frac{\lamh}{4}\chi^4,
\eeq
with $g^2>0$.  The significance of such a term for \pr was studied in 
lattice simulations \cite{pr97} and analytically \cite{ac97}, 
but in the absence of metric perturbations.  
Bassett and Viniegra \cite{bv99} included metric perturbations, 
but ignored backreaction.  Essentially, for $\lamh\gtrsim\lam$ 
we expect the $\chi$ self-interaction to limit the growth of 
perturbations as compared with the $\lamh=0$ case studied above, 
due to the presence of the ``potential wall'' $(\lamh/4)\chi^4$.

   More precisely, the linearized equation of motion for the $\chi$ field 
perturbation becomes, with $\chi$ self-interaction but ignoring metric 
perturbations,
\bea
\ddot{\delchi}_k + 3H\delchid_k &+& \left({k^2\over a^2} + 3\lamh\chi^2
  + g^2\phi^2\right)\delchi_k \nonumber\\
                                &+& 2g^2\phi\chi\delphi_k = 0.
\label{delchisi}
\eea
Thus for small enough initial $\chi$ background, the initial behaviour of 
the modes will be essentially unchanged from the $\lamh=0$ case.  
However, when the $\chi$ background grows to the point that 
$\chi^2/\phi^2\gtrsim g^2/\lamh$, the analytical \pr theory of Section 
\ref{linanal} no longer applies, and we may expect the perturbations 
to stop growing.  Since, as discussed above, for the significant 
production of \sh modes we require $g^2\simeq\lam$, we expect that 
$\chi$ self-interaction will shut down the resonance when 
$\chi^2/\phi^2\sim \lam/\lamh$, as long as $\lamh\gtrsim\lam$.  
If $\lamh<\lam$, then the $\chi^4$ interaction term will not lead to a 
shutdown of the resonance since (based on our numerical simulations) 
the homogeneous $\chi$ field never substantially exceeds the value of the 
inflaton background.

\begin{figure}
%\centerline{\psfig{figure=/home/zibin/ph/paper/fig5.ps,height=8cm}}
\centerline{\psfig{figure=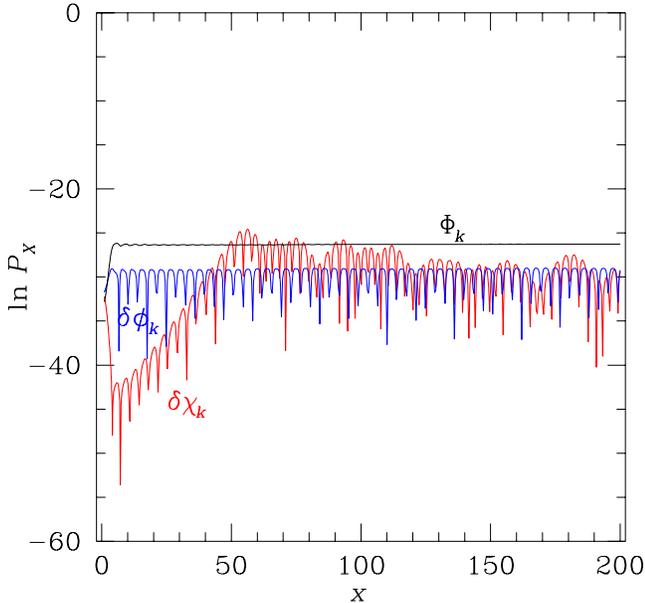,height=8.7cm}}
\caption{Same as for Fig. \ref{br}, except with $\chi$ field 
self-coupling $\lamh=10^{-10}$.  The $\chi$ self-coupling causes the 
early termination of growth.}
\label{si}
\end{figure}

   We have confirmed this expectation numerically, and we give an example 
of our results in Fig. \ref{si}.  Here we have included Hartree 
backreaction and metric perturbations, and used coupling constant values 
$\lam=10^{-14}$, $g^2/\lam=2$, and $\lamh=10^{-10}$, and initial 
backgrounds $\phi(t_0)=4\mpl$ and $\chi(t_0)=10^{-6}\mpl$.  We 
indeed observe the termination of the \sh modes' growth at approximately 
the time when $\chi^2/\phi^2 = \lam/\lamh$.

   Note that, {\em in the absence of backreaction,} Bassett and 
Viniegra observed a continued slow growth of \sh perturbations 
after the initial termination of the resonance when 
$\chi^2/\phi^2\sim \lam/\lamh$ \cite{bv99}.  We confirmed 
this result; however, we note that when we include the 
backreaction term $3\lamh\flchi2$ in the evolution equations, we 
expect backreaction to become important also at the time that 
$\chi^2/\phi^2\sim \lam/\lamh$, with our choice $\chi^2\simeq\flchi2$.  
Hence, as seen in Fig. \ref{si}, the slow growth is completely suppressed.

\subsection{Negative coupling}

   The presence of $\chi$ self-coupling means that we no longer 
require $g^2\geq0$ for global stability.  In fact, for the case $g^2<0$, 
the potential will be bounded from below for $\lam\lamh/g^4>1$ 
\cite{gpr97}.  This {\em negative coupling} case was studied in the 
absence of metric perturbations using lattice simulations in \cite{gpr97}, 
and without backreaction in \cite{fb00}.  The behaviour of the 
fields is qualitatively different in the negative and positive 
coupling cases.  For $g^2<0$, potential minima exist with non-zero 
homogeneous part of the $\chi$ field.  Thus, assuming the fields 
fall into these minima, the problem of choice of $\chi$ background 
discussed in previous sections for the positive coupling case is 
alleviated.

   For initial homogeneous $\chi$ fields large enough ($\chi(t_0)\gtrsim\mpl$), 
we find numerically that the fields fall into the potential minimum by 
the end of inflation, and the two fields subsequently evolve in step 
during preheating.  This effectively reduces the system to a 
single-field system, and hence no resonance is possible on \sh scales.

   To see this explicitly consider the case $\lamh=\lam$, for which the 
symmetry of the potential requires the potential minima to lie along 
$\chi^2=\phi^2$.  If we choose the same initial signs for $\chi$ and $\phi$, 
then during \ph the backgrounds lie along the attractor $\phi=\chi$.  Similarly, 
since the behaviour of \sh modes is essentially the same as that of 
the backgrounds, we have $\delchi=\delphi$ during preheating.  
Then the perturbation equation (\ref{delchisi}) becomes
\beq
\ddot{\delchi}_k + 3H\delchid_k + \left[{k^2\over a^2}
 + 3(\lam+g^2)\phi^2\right]\delchi_k = 0.
%\label{delchisi}
\eeq
Thus the effective mass of the $\delchi$ oscillations is precisely three 
times the effective mass of the background inflaton oscillations ({\it cf } 
\eq(\ref{kg})), so that just as with the case of the inflaton 
perturbations in \eq(\ref{delphiconf}), there will be no resonance 
on \sh scales for all allowed values of $g^2$.  We have confirmed this 
numerically; indeed more generally, as long as initially $\chi(t_0)\sim\mpl$ 
but for any $\lamh\geq\lam$, the two fields will be proportional during 
\ph and no \sh resonance will result.

  This result assumes that during \ph only the ``field'' $\delchi+\delphi$ 
is excited.  If orthogonal field excitations $\delchi-\delphi$ are 
present, they can grow resonantly.  The effective squared mass of 
$\delchi-\delphi$ excitations is $(3\lam-g^2)\phi^2$, so that according 
to the analytical \pr theory of Section \ref{linanal}, \sh resonance 
will occur near $3\lam-g^2=2n^2(\lam+g^2)$, for integral $n$ (we require 
$n\geq2$ for negative $g^2$).  That is, \sh $\delchi-\delphi$ modes will 
grow for $g^2\simeq\lam(3-2n^2)/(1+2n^2)$.  However, numerically we 
observe only extremely small components $\delchi-\delphi$ by the end of 
inflation, so their growth is substantially delayed.

   On the other hand, for small initial homogeneous part $\chi(t_0)\ll\mpl$, 
we find that the potential minima are not reached by the end of inflation, 
and the two fields evolve in a very complicated manner during preheating.  
The analytical theory of \pr cannot be applied, but numerically we 
do find roughly exponential growth of \sh modes in this case, as found 
in \cite{fb00}.  The growth rate increases as $g^2$ decreases towards 
the value at which global instability sets in, $g^2=-\sqrt{\lam\lamh}$.

\begin{figure}
%\centerline{\psfig{figure=/home/zibin/ph/paper/fig6.ps,height=8cm}}
\centerline{\psfig{figure=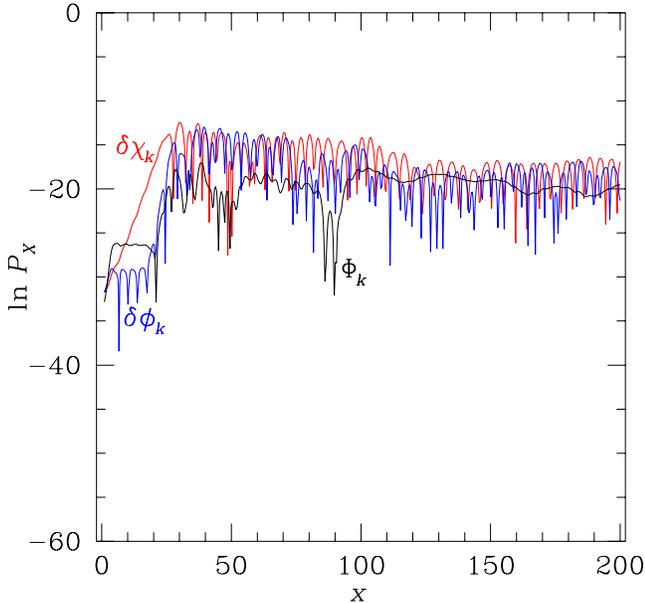,height=8.7cm}}
\caption{Same as for Fig. \ref{si}, except for the negative coupling 
case, with parameters $\lam=10^{-14}$, $g^2=-0.5\lam$, and $\lamh=\lam$.  
Here the growth is comparable to the $\lamh=0$ case of Fig. \ref{br}, 
even though analytical \pr theory cannot be applied here.}
\label{nc}
\end{figure}

  We have illustrated this case in Fig. \ref{nc}, using the parameter 
values $\lam=10^{-14}$, $g^2=-0.5\lam$, $\lamh=\lam$, $\phi(t_0)=4\mpl$, 
and $\chi(t_0)=10^{-6}\mpl$.  Here the growth rates and final power 
spectra values are comparable to the $\lamh=0$ case of Fig. \ref{br}, 
though the $\delchi$ field is not damped during 
inflation for negative coupling.  For $\lamh>\lam$, the growth is 
terminated early, just as in the positive coupling case.

\section{Summary and Discussion}

   In this paper we have studied backreaction effects on the growth of 
super-Hubble cosmological fluctuations in a specific class of two field 
models with a massless inflaton $\phi$ coupled to a scalar field $\chi$.  
Our study was based on the Hartree approximation.

   For the non-self-coupled $\chi$ field case, we found that 
backreaction has a crucial effect in determining the final 
amplitude of fluctuations after preheating.  For values of the 
coupling constants satisfying  $g^2/\lam=2$ (the ratio predicted in 
supersymmetric models), the predicted amplitude of the \sh metric 
perturbations at the end of preheating is too large to be consistent 
with the COBE normalization, thus apparently ruling out such models.  
One possible loophole is the \br contribution from super-cosmological 
scale fluctuations.  For sufficiently long periods of inflation, the 
predicted amplitude can be consistent with the COBE normalization.  
In addition, the final amplitude of the fluctuation spectrum is 
independent of the coupling constant $\lambda$.  
Note that the growth of inflaton fluctuations $\delphi_k$ (and hence 
metric perturbations $\Phi_k$) occurs in these models either through 
coupling to $\delchi_k$ via a homogeneous background $\chi$ field or 
through nonlinear evolution effects.

   The situation for $g^2/\lam \gg 1$ is very similar to the previously 
studied case of a massive inflaton in the broad resonance regime 
\cite{i99,js99,llmw99}.  Cosmological-scale 
$\delchi_k$ modes are significantly damped during inflation, and 
the end of resonant growth is determined by the growing small-scale 
modes.  Already for the second resonance band (centred at $g^2/\lam=8$) 
cosmological metric perturbations are not amplified above the 
COBE normalization value.  This implies that \ph does not alter the 
standard predictions for the $\Phi_k$ normalization in $(\lam/4)\phi^4$ 
inflation for the second and all higher resonance bands.  The important 
difference between the model we have studied and the massive inflaton 
case is that, in the massive model, weak \sh suppression at small 
$g^2$ is accompanied by weak resonant growth during preheating 
\cite{js99}, so that no significant \sh amplification is possible.

   The inclusion of $\chi$ field self-interaction alters the evolution 
in a predictable way: the resonant growth stops when 
$\chi^2/\phi^2\sim \lam/\lamh$, as long as $\lamh\gtrsim\lam$.  
This means that we are unable to rule out models (on the basis of a 
too large production of metric perturbations) with 
$\lamh/\lam\gtrsim10^{4}$.  In the negative coupling case, there are 
two possibilities.  For large initial $\chi$ backgrounds, 
$\chi(t_0)\sim\mpl$, the system becomes essentially single-field, 
and no resonance occurs (at least until late times).  For small 
initial $\chi$, exponential growth occurs for large enough allowed $|g^2|$.

   The Hartree approximation provides a useful approach for the 
inclusion of the effects of backreaction.  However, 
as mentioned in Section III, this approximation misses 
terms which could contribute to the evolution of fluctuations in an 
important way.  We believe, nevertheless, that our results are 
sufficiently accurate to predict, for 
the models studied, whether or not the metric perturbation 
amplitude after preheating is consistent with the COBE measurement.  
Nonlinear effects, or rescattering, will primarily affect 
the detailed evolution of matter fields after \br is important.  
Still, it is of great interest to extend our analysis 
to a full nonlinear treatment, as was done in the absence of 
gravitational fluctuations in \cite{kt96,kt,pr97}, and including 
metric fluctuations in \cite{pe} for single field models.
%Note that the full nonlinear analysis including metric fluctuations has to 
%deal with the full complexities of nonlinear gravity.  In \cite{pe99} the 
%problem was simplified by assuming planar symmetry.  In light of the 
%limitations of the Hartree approximation, our conclusions 
%concerning the compatibility of the models studied with the current 
%observational constraints must be viewed as tentative.

%[req'd g large for massive infl case.]
%[expect $g^2 \simeq 10^{-2}$, though, except for susy?]
%[$\xi < 0 \Rightarrow$ growth even for lg $g^2/\lam$?]

%[growth of $\delphi$ and $\Phi$ due to coupling with $\delchi$ 
%through background $\chi$ or backreaction?, contrary to \cite{bv99},
%unimportance of background $\chi$?]

%[size of $\Phi$ power spectrum after preheating compared with COBE, 
%discussion of $\lam$ dependence...]

%[nonlinear results approx?]

\acknowledgements
We wish to thank F. Finelli for useful discussions.
This research was supported by the Natural Sciences and Engineering 
Research Council of Canada. The work of R.B. was supported in part by
the U.S. Department of Energy under Contract DE-FG02-91ER40688, TASK A.

%\bibliography{bib}

\end{document}